\documentclass[reprint,aip,amsmath,amssymb,nofootinbib]{revtex4-2}
\usepackage{amssymb}
\usepackage{amssymb}
\usepackage{amssymb}
\usepackage{graphicx}
\usepackage{graphics}
\usepackage{amsmath}
\usepackage{placeins}
\usepackage{hyperref}
\usepackage[dvipsnames]{xcolor}
\usepackage{wasysym}
\usepackage{soul}
\usepackage{float}
\usepackage{array}
\usepackage{tabulary}
\usepackage{caption}
\usepackage{fancyhdr}
\usepackage{wrapfig}

\usepackage{tikz}
\usetikzlibrary{shapes.geometric, arrows}

\usepackage[normalem]{ulem}
\usepackage{graphicx}
\usepackage{color,soul}
\usepackage{subfigure}
\usepackage{dcolumn}
\usepackage{bm}
\usepackage{hyperref}
\usepackage{url}
\usepackage{multirow}
\hypersetup{
    colorlinks=true,
    linkcolor=blue,
    filecolor=blue,      
    urlcolor=blue,
    citecolor=blue,
}

\usepackage[makeroom]{cancel}
\begin{document}

\title{Travelling Wakefield Tube: THz Source Powered by Nonrelativistic Electron Beam}

     \author{\firstname{Mitchell E.} \surname{Schneider}}
	\email{schne525@msu.edu}
    \affiliation{Department of Electrical and Computer Engineering, Michigan State University, MI 48824, USA}
    \affiliation{Department of Physics and Astronomy, Michigan State University, East Lansing, MI 48824, USA}
    \affiliation{Accelerator Operations and Technology Division, Los Alamos National Laboratory, Los Alamos, NM 87545, USA}
    \author{\firstname{Emily} \surname{Jevarjian}}
    \affiliation{Department of Physics and Astronomy, Michigan State University, East Lansing, MI 48824, USA}
    \author{\firstname{Ben} \surname{Sims}}
    \affiliation{Department of Electrical and Computer Engineering, Michigan State University, MI 48824, USA}
    \affiliation{Department of Physics and Astronomy, Michigan State University, East Lansing, MI 48824, USA}
   \author{\firstname{Alexander} \surname{Altmark}}
    \affiliation{Department of Physics, Saint-Petersburg Electrotechnical University “LETI”, 197376 Saint Petersburg, Russian Federation}
    \author{\firstname{Bas} \surname{van der Geer}}
    \affiliation{Pulsar Physics, 5614 BC Eindhoven, The Netherlands}
    \author{\firstname{Sergey V.} \surname{Baryshev}}
	\email{serbar@msu.edu}
	\affiliation{Department of Electrical and Computer Engineering, Michigan State University, MI 48824, USA}

\begin{abstract}
High peak power, tunable, narrowband terahertz emitters are becoming sought after given their portability, efficiency, and ability to be deployed in the field for industrial, medical, and military applications. The use of accelerator systems producing THz frequencies via Cherenkov radiation, generated by passing an electron beam through a slow-wave wakefield structure, is a promising method to meet future THz requirements. To date, efforts have been dedicated to analysis and design of sources utilizing laser seeded bunched electron beam drivers with relativistic energies beyond 5 MeV. Presented here is a wakefield THz generation scheme based on passing a long quasi-dc nonrelativistic beam (200 keV) through a dielectric loaded travelling wave structure. Reduced energy allows for compactness and portability of the accelerator as the size and weight of the dielectric slow wave structure is vanishingly small compared to the accelerator unit. The presented scheme can serve as a tunable high peak power THz source operated between 0.4-1.6 THz and produces power gain by a factor of five with an average efficiency of 6.8\%.
\end{abstract}

\maketitle

\section{Introduction}\label{S:intro}
The THz frequency range has a unique niche in nondestructive imaging, remote sensing, biological imaging, and communication\cite{1,2,3,4,5,6,7,8,9}. Expanding or improving these applications is reliant upon developing high pulsed peak power THz sources that are tunable \cite{1}. Practical table-top THz sources that are currently available are optically pumped semiconductor crystals, such as ZnTe, where fs pulses are converted into THz signal, or electrically driven quantum cascade lasers (QCLs). Another class of devices called TeraFET exists\cite{10} where resonant or broadband THz generation is possible. These devices however suffer either from low power or low operating frequencies in sub 1 THz, or necessity to be operated under a cryogenic temperature, or being broadband or non-tunable.

There is a class of applications where generation of 100 to 1,000 GHz narrow band signals are of particular interest. For instance, electron beam driven THz oscillators and amplifiers could support the ongoing development of compact accelerators operating in W-band\cite{11}, various environments monitoring in the 200-300 GHz window, time-resolved pump-probe experiments, and many others. In all such applications, simplicity, compactness, and cost are prioritized. THz vacuum electronic devices (VEDs) are slow wave devices that include traveling wave tube (TWT) or backward wave oscillators (BWO). TWT-like systems produce high peak power radiation in the sub-THz regime, from MHz to W-band. For example, at frequencies of 0.2 THz, the peak power can achieve $\sim$10 W; however, as the frequency increases, it quickly drops to $\sim$0.5 W at $\sim$0.8 THz. To the authors’ knowledge, TWTs and other VEDS have not yet been reported in the literature to be capable of operating at or over 1 THz\cite{3,4}.

High operating frequencies near 1 THz can be achieved while maintaining peak power by generating an electromagnetic signal in the form of either synchrotron, Cherenkov, or Smith-Purcell\cite{12} radiation released using linacs. Synchrotron sources are large-scale facilities and cannot be deployed in the field. In that sense, linac systems are preferable as they allow for simpler modular system design. Like TWTs, to enable coherent/narrow band THz output, linac-based generators require an initial $\sim$100 fs/ps pulse which seeds the electron beam. The beam subsequently acquires sub-100 fs modulation which is further converted into THz frequencies when the bunched beam passes through a slow wave structure. In linac systems that combine acceleration and drift sections, the seed signal is the fs laser or a special type of collimating/rotation techniques. These techniques can be combined using a mask and a pair of dipole magnets (the dogleg) or a pair of deflecting cavities to make the THz radiation\cite{13,14}. In synchrotrons, seeding is done using the undulator\cite{15,16}. In linac systems, in the ultrarelativistic ($\gtrsim$10 MeV)\cite{17} or nonrelativistic ($\sim$1 keV)\cite{18} energy range, the effect of seeding can be further amplified through the space charge force. Again, ultra-relativistic sources require a large-scale accelerator facility which means that these sources cannot be used for in the field applications. While the non-relativistic $\sim$1 keV sources need a significant amount of charge on the order of a few nano coulombs, which is hard to produce with commonly available sources for these devices.

External seeding, however, can be avoided. It is sufficient to send a long electron pulse through a slow wave wakefield (disk loaded/corrugated metal or dielectric) structure. If the Cherenkov condition is satisfied, the head of the electron pulse excites waveguide modes via the wakefield mechanism, in extreme cases causing beam breakup instability. Wakefield wave packets travel forward while unfolding backwards and impose energy modulation on the rest (tail) of the pulse. If the beam drifted long enough, spatial ballistic bunching\cite{19} will take place because different parts of the bunch travel with different velocities. It takes a fair amount of distance (a few meters) spatial ballistic bunching to occur for MeV electrons. To promote bunching (i.e. to happen at a shorter length scale), a four-dipole prism (chicane) can be used\cite{20}. Its strength can be tuned, allowing for tunability of the bunching distance. When the linac beam (bunched using either laser or head-tail wakefield interaction) is sent through a second slow wave waveguide, high peak power narrowband THz frequencies are generated via the same Cherenkov mechanism. This method usually results in terahertz peak energy on the order of $\sim$10 $\mu$J\cite{21} and can reach up to 2 THz \cite{22}. Even successful, all proof-of-concept demonstrations were done in the MeV energy range. Relativistic MeV systems are very large and complex and additionally produce unsafe x-ray/$\gamma$-radiation which requires an exceptional amount of shielding, adding more weight and cost.

Proposed here is a travelling Cherenkov wakefield generator where high power narrow band THz signals are generated using a nonrelativistic 200 keV beam (Lorentz factor $\gamma$$\sim$2). For this purpose, a long electron pulse (100 ps) is transmitted through a quartz lined slow wave cylindrical waveguide. Head-tail action modulates the bunch and microbunching then emerges only a cm away from the waveguide which is a result of strong space charge affecting the nonrelativistic beam. Analytically and computationally, it is shown that such a method leads to very efficient high peak power THz generation. The proposed concept offers a practical path for designing a simple table-top high peak power tunable THz source.

\section{Computational methodology}\label{S:comp}
\subsection{Self-imposed wakefield in cylindrical DLW: First order monoenergetic estimations}\label{S:mono}

\begin{figure}
	\centering\includegraphics[width=8cm]{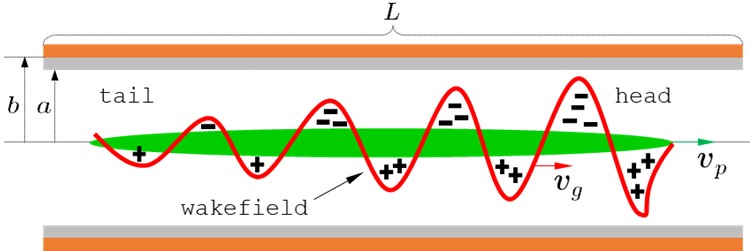}
	\caption{Diagram showing the evolution of a self-generated wakefield inside of a cylindrical DLW. Here, $a$ is the internal radius of the of vacuum channel, $b$ is the outer radius of the structure, and $L$ is the length. The dielectric (grey layer) for this DLW is quartz ($\varepsilon_r$=3.86), and the outer wall (orange) is coated copper.}\label{F:1}
\end{figure}

As the head of an electron beam enters a Dielectrically Lined Waveguide (DLW, see Fig.~\ref{F:1}), it produces a Cherenkov wakefield potential inside of the DLW that, in turn, acts on the tail of the beam. This is called head-tail self-induced interaction, and is schematically diagrammed in Fig.~\ref{F:1}. This Cherenkov radiation packet unfolds backwards and travels forward at a group velocity ($v_g$) in this quartz-lined slow-wave cylindrical waveguide. The wakefield-beam interaction only takes place inside of the DLW. For useful oscillator design, the DLW must be sufficiently long for the slow-wave wakefield to propagate backwards, reach, and modulate the entire tail of the beam. This is called the interaction length and is directly proportional to $v_g$.

Firstly, the group velocity must be calculated. For a cylindrical DLW, it must first be considered that the beam head will excite the TM$_\text{0m}$ modes. The longitudinal wakefield Green’s function, which is the solution of the wave equation containing a delta function which represents the head of the beam, is of the form

\begin{equation}\label{E:1}
E_z^0(r,\theta,\zeta)=-4Q \sum_{n=1}^{\infty}\Phi^0_{E_z}(0,k_\text{0,m})\text{cos}(k_\text{0,m},\zeta),
\end{equation}
where $\zeta=z-vt$ is the distance behind the bunch, $\Phi^0_{E_z}(v,k)$ is the amplitude of the wakefield, and $k_\text{0,m}$ are the roots for the dispersion function for the waveguide and can be derived using a methodology outlined in Ref.\onlinecite{23}. The one-dimensional wakefield profiles were calculated using the code “Waveguide”\cite{24} which finds and exports a Green’s function using specified geometry and boundary conditions. It also numerically determines the wave number and frequencies of all of the TM0m modes. The minimal length of the DLW can thus be found from the group velocity for each TM0m mode. The group velocity relative to the speed of light ($\beta_g=\frac{v_g}{c}$) is found using the zeros of the Bessel function ($Z_\text{0m}$) of the first kind which analytically describe TM$_\text{0m}$ modes (as obtained from solving the Maxwell’s equations) as

\begin{equation}\label{E:2}
Z_{0m}=Zeros[J_0(k(\beta_g)b)]_m.
\end{equation}

Then, the group velocity for each dominant mode present in the DLW is calculated as

\begin{equation}\label{E:3}
\left\{
\begin{aligned}
k(\omega,\beta_g)&=\sqrt{(k^2-k_z^2)}=\frac{\omega}{c}\sqrt{\varepsilon-\frac{1}{\beta^2_g}}\\
k&=\frac{\omega\sqrt{\varepsilon}}{c}\\
k_z&=\frac{\omega}{\beta_gc}\\
\beta_{gnm}&=\left(\varepsilon-\left(\frac{z_{nm} c}{b\omega_{nm}}\right)^2\right)^{-\frac{1}{2}}
\end{aligned}
\right.
\end{equation}
Finally, the minimal required DLW length that would guarantee the entire beam tail efficiently modulated by the main TM$_\text{01}$ mode is found as
\begin{equation}\label{E:4}
L=\beta_{g01}\cdot c\cdot PL,
\end{equation}
where $PL$ is the pulse length in units of time and $c$ is the speed of light. Eq.~\ref{E:4} yielded that the minimal DLW length is 1.36 cm for $\beta_{g01}$=0.9 at 200 keV. This analytical estimation was confirmed by simulations using a convolution algorithm-based module (ver. 3.40) in General Particle Tracer (GPT) in 3D. The length was rounded to 1.4 cm to accommodate additional longitudinal beam expansion due to space charge. The DLW cross section design was optimized in terms of $a$ and $b$ for the most efficient energy beam-to-radiation conversion for the TM$_\text{01}$ mode. The higher harmonics (up to TM$_\text{05}$) are in the THz regime and can be extracted downstream as a result of the space charge and wakefield forces’ interplay. The entire set of optimized geometry and mode parameters are listed in Tables~\ref{T:1} and~\ref{T:2}.

\begin{table}[!]
\centering
\caption{DLW and beam parameters}
\begin{tabular}{ l  c | c  c}

Parameter        &  & Value \\ 
\hline
\hline
$a$              &  & 0.36 mm \\
$b$              &  & 0.25 mm \\
$L$              &  & 1.4 cm \\
$\varepsilon_r$  &  & 3.86 \\
N of electrons  &  & $10^6$ \\
Energy           &  & 200 keV  \\
$R_{beam}$       &  & 50 $\mu$m \\
$PL$             &  & 50 ps \\
$Q$              &  & 20 pC
\end{tabular}
\label{T:1}
\end{table}

\begin{table}[!]
\centering
\caption{Major TM modes for the given DLW geometry and beam parameters}
\begin{tabular}{ l  c | c  c}

Mode             &  & Frequency (THz) \\ 
\hline
\hline
TM$_\text{01}$  &  & 0.470 \\
TM$_\text{02}$  &  & 1.442 \\
TM$_\text{03}$  &  & 2.451 \\
TM$_\text{04}$  &  & 3.466 \\
TM$_\text{05}$  &  & 4.482
\end{tabular}
\label{T:2}
\end{table} 

\begin{figure}
	\centering
	\includegraphics[width=8cm]{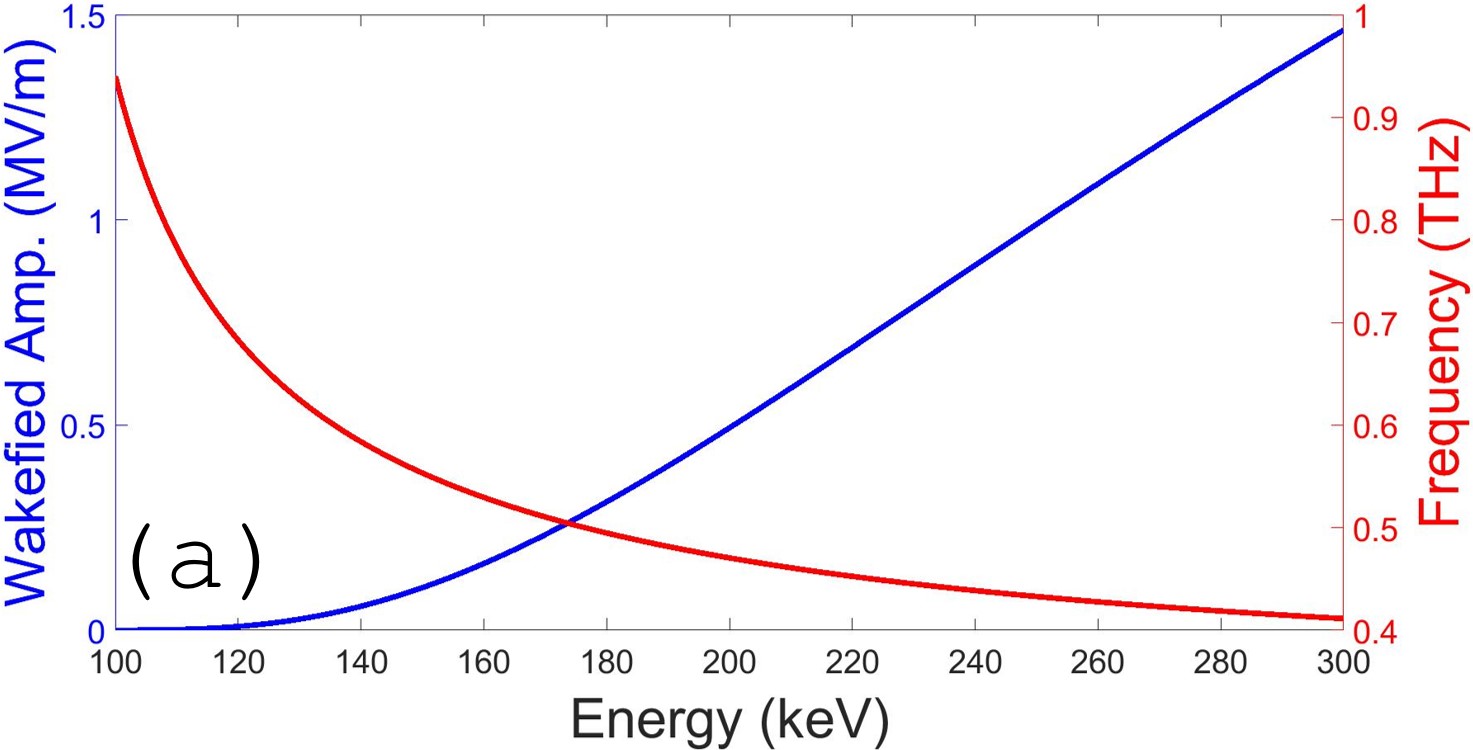}
	\hspace*{-0.8cm}\includegraphics[width=7.7cm]{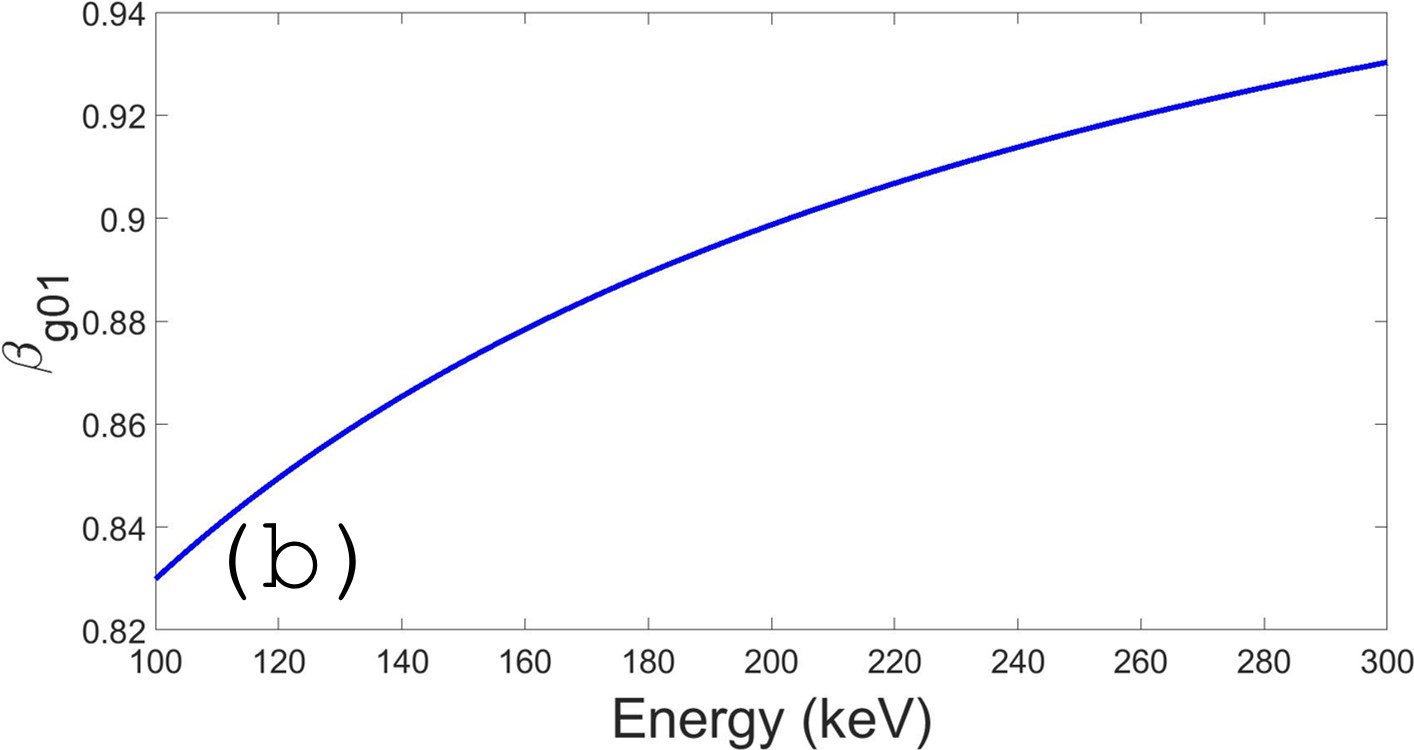}
	\caption{(a) The frequency and wakefield dependence of energy for the geometry presented in Table~\ref{T:1}. (b) The energy dependence for the group velocity for the TM$_\text{01}$ mode in the laboratory frame as Eq.~\ref{E:3} is in a moving reference frame when the wakefield is generated. To find the group velocity needed to calculate the minimum length of the DLW, a Lorentz velocity transformation is performed on Eq.~\ref{E:3}.}\label{F:2}
\end{figure}

The finalized geometry allows for the estimation of corresponding energy loss in the DLW as

\begin{equation}\label{E:5}
E_{loss}=\frac{E_z^0\cdot v_{g01}\cdot P L}{2\cdot E_k},
\end{equation}
where $E_z^0$ is the amplitude of the Green’s function, $E_k$ is the kinetic energy of the beam, and $v_{g01}$ is the group velocity for the TM01 mode in the lab frame. For initial energy $E_k$=200 keV, it was found that the beam experiences a 10\% loss of its energy. Such a loss is comparable with the initial beam energy and unlike in relativistic case cannot be ignored: wake amplitude and frequency depend nonlinearly on the beam energy in sub-MeV range. To elaborate on this value and find the best practical beam energy, the wake amplitude, frequency, and group velocity were determined as functions of the beam energy between 100 and 300 keV. Energies between 100 and 150 keV are ruled out as the wake amplitude non-linearly vanishes and therefore those energies do not have practical implications. Energies above 200 keV are known to produce nonlinearly enhanced x-rays and are not practical for a tabletop experimental setting as they require excessive radiation shielding. The 10\% loss atop the initial energy of 200 keV indicates that the energy range of interest is between 180 and 200 keV. From Fig.~\ref{F:2}, it can be seen that all of the quantities of interest behave linearly but do change within a significant range that cannot be neglected and needs to be dealt with carefully. To account for the energy loss, the corresponding wakefield retuning, and obtaining realistic oscillator operation, we present a methodology called polyenergetic Green’s function method in the next section and demonstrate its self-consistent implementation in GPT.

\subsection{Calculations of high-power THz radiation in a DLW: Second order polyenergetic Green’s function}\label{poly}

\begin{figure}[!]
\begin{tikzpicture}[node distance=1.5cm]
\tikzstyle{startstop} = [rectangle, minimum width=3cm, minimum height=1cm,text centered, draw=black, fill=green!30]
\tikzstyle{arrow} = [->,>=stealth]
\centering
\node (in1) [startstop] {2D longitudinal phase space from GPT};
\node (in2) [startstop, below of=in1] {Phase space density $\Phi(z,E)$};
\node (in3) [startstop, below of=in2] {$\rho_y=Q\frac{\int^{\infty}_{-\infty}\Phi(z,E)dE}{\int^{\infty}_{-\infty}\int^{\infty}_{-\infty}\Phi(z,E)dEdz}$};
\node (in4) [startstop, below of=in3] {$I\left(\frac{z}{v}\right)=v(z)\cdot \rho_y(z)\cdot\pi\cdot r^2$};
\node (in5) [startstop, below of=in4] {$P(f)=[FFT(I(t)]^2\cdot R_s(f)$};
\draw [arrow] (in1) -- (in2);
\draw [arrow] (in2) -- (in3);
\draw [arrow] (in3) -- (in4);
\draw [arrow] (in4) -- (in5);
\end{tikzpicture}
\caption{Data processing pipeline using the longitudinal phase space extracted from GPT that results in the frequency spectrum where $\rho_y$ is the charge density, $P(f)$ is the power frequency spectrum in units of Watts, and $R_s$ is the surface resistance assuming that the conductivity of the walls.}\label{F:3}
\end{figure}
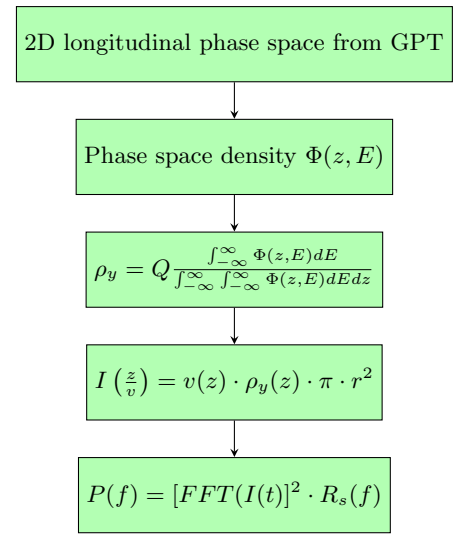

To account for the energy loss related effects highlighted in Fig.~\ref{F:2}, a new version of the convolution-based wakefield module in GPT was developed which subdivided the phase space into energy bands. GPT treats each of the energy individually, allowing for each band to have a separate Green’s function, when combined, representing a polyenergetic Green’s function. A library of Green’s functions generated with “Waveguide” were imported into GPT’s new polyenergetic wakefield module. To translate the longitudinal phase space extracted from GPT into a frequency spectrum, we used a data processing pipeline as outlined in Fig.~\ref{F:3}. This procedure was originally described in a paper by Lemery and Piot\cite{25}.

The convolution in GPT with different Green’s functions corresponding to different energy bands was done continuously and self-consistently as the particles traveled through the DLW. It was concluded that the frequency (TM$_\text{01}$) and the amplitude of the Green’s function should have only a 1\% difference between each energy band. As Fig.~\ref{F:2}a shows, the frequency depends exponentially on the energy; therefore, energies less than 150 keV would cause an undesired wideband output frequency. For energies greater than 150 keV, the linear dependency in the amplitude is the dominating factor when determining the width of the energy band. Using this linear dependency, it was found that the width of the energy band for a 1\% difference should be 0.5 keV.

\section{Results from polyenergetic Green’s function method, discussion and benchmarking against CST PIC}\label{results}

\begin{figure*}[!]
	\centering
	\includegraphics[height=4cm]{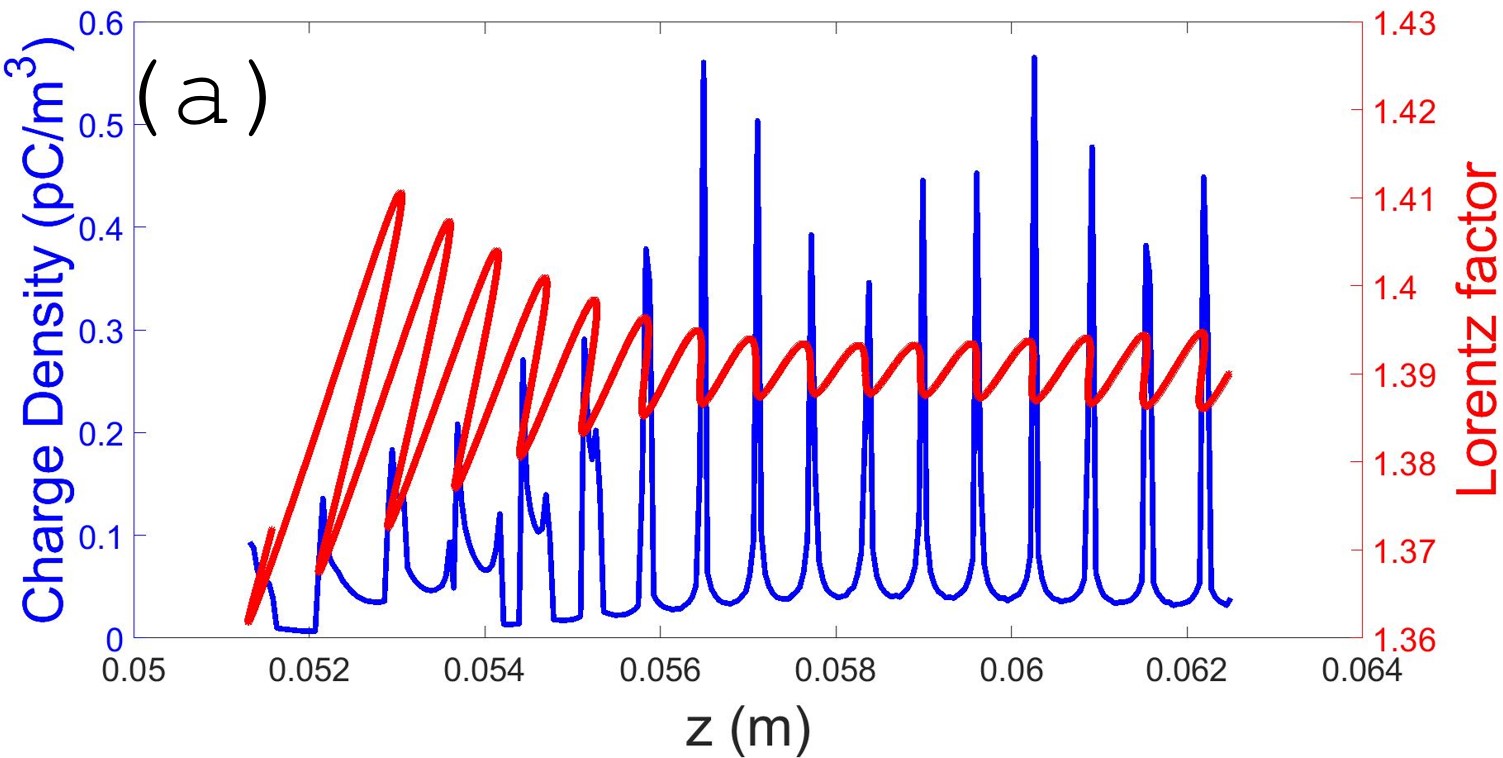}\hspace{0.5cm}\includegraphics[height=4cm]{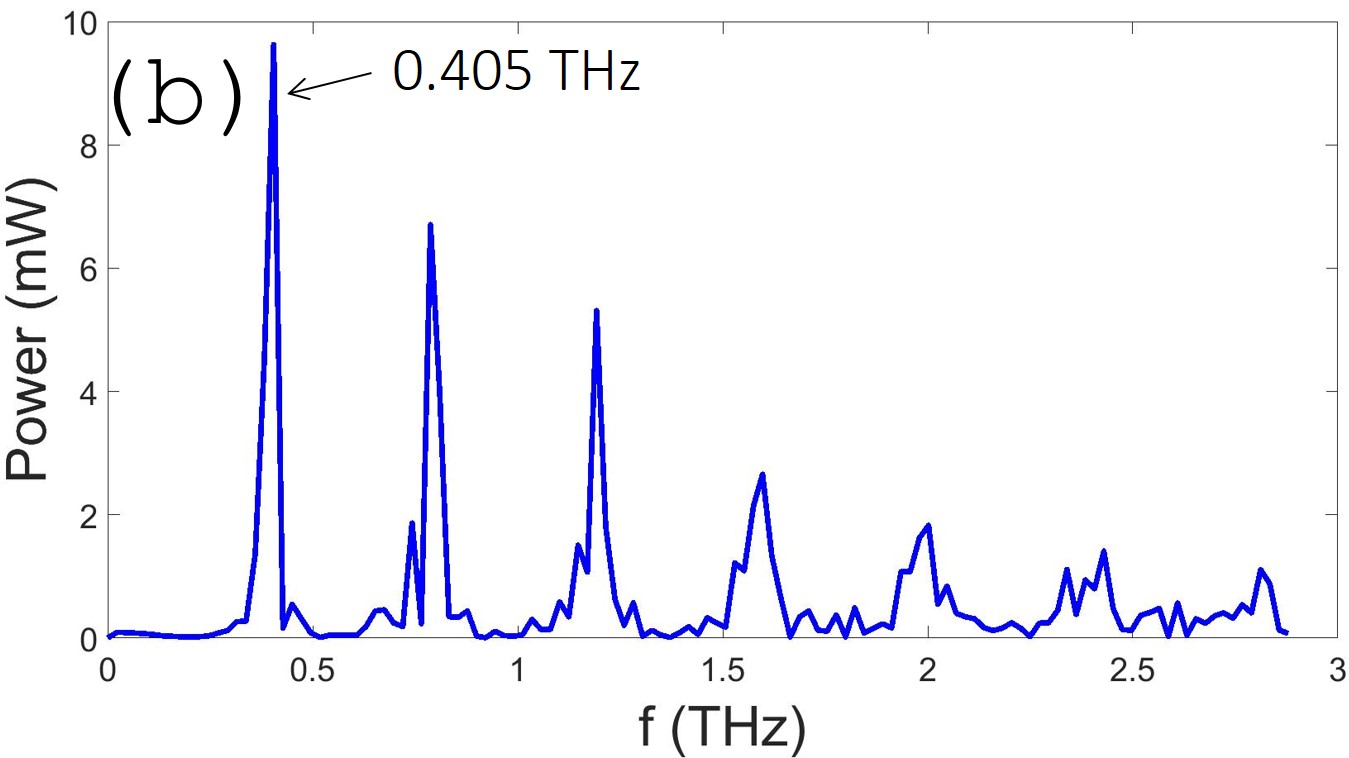}
	\includegraphics[height=4cm]{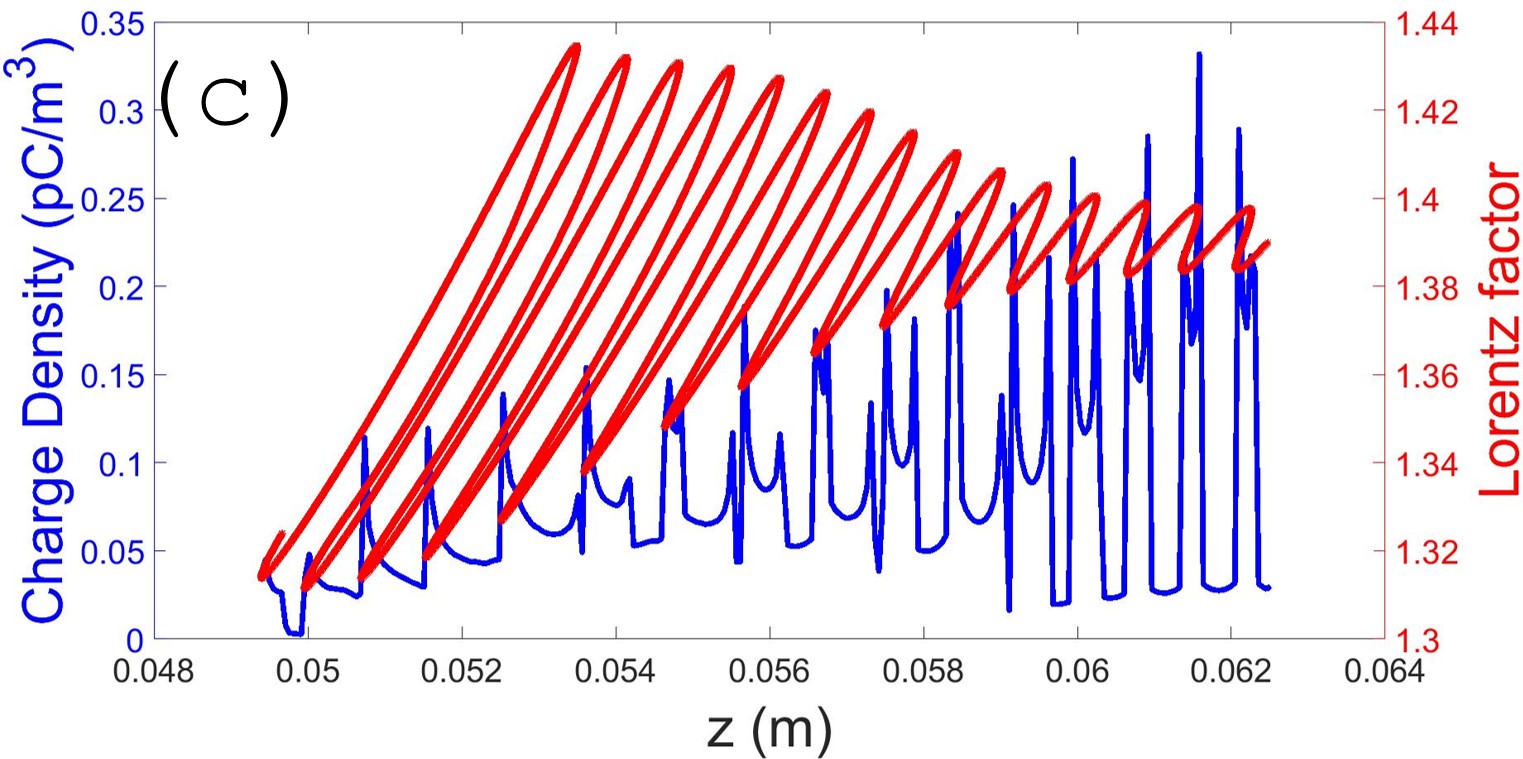}\hspace{0.5cm}\includegraphics[height=4cm]{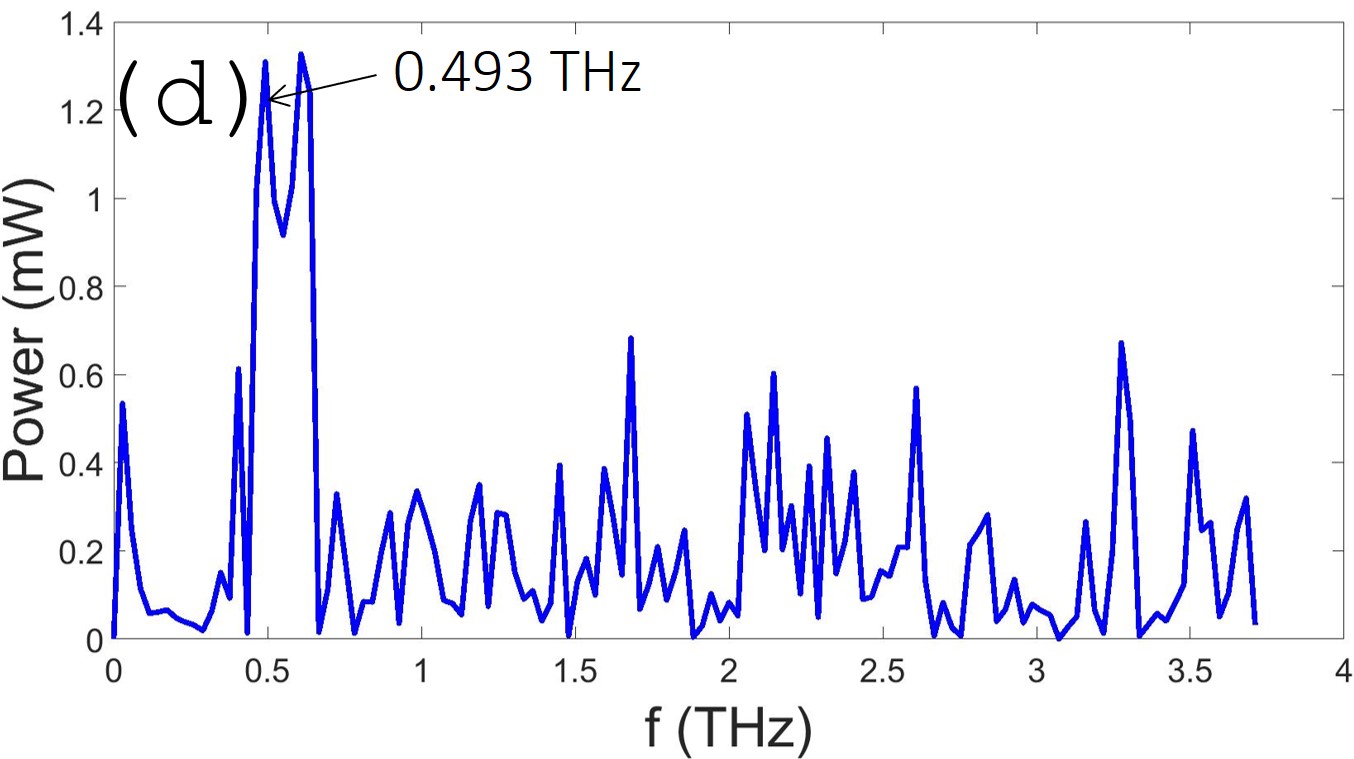}
	\includegraphics[height=4cm]{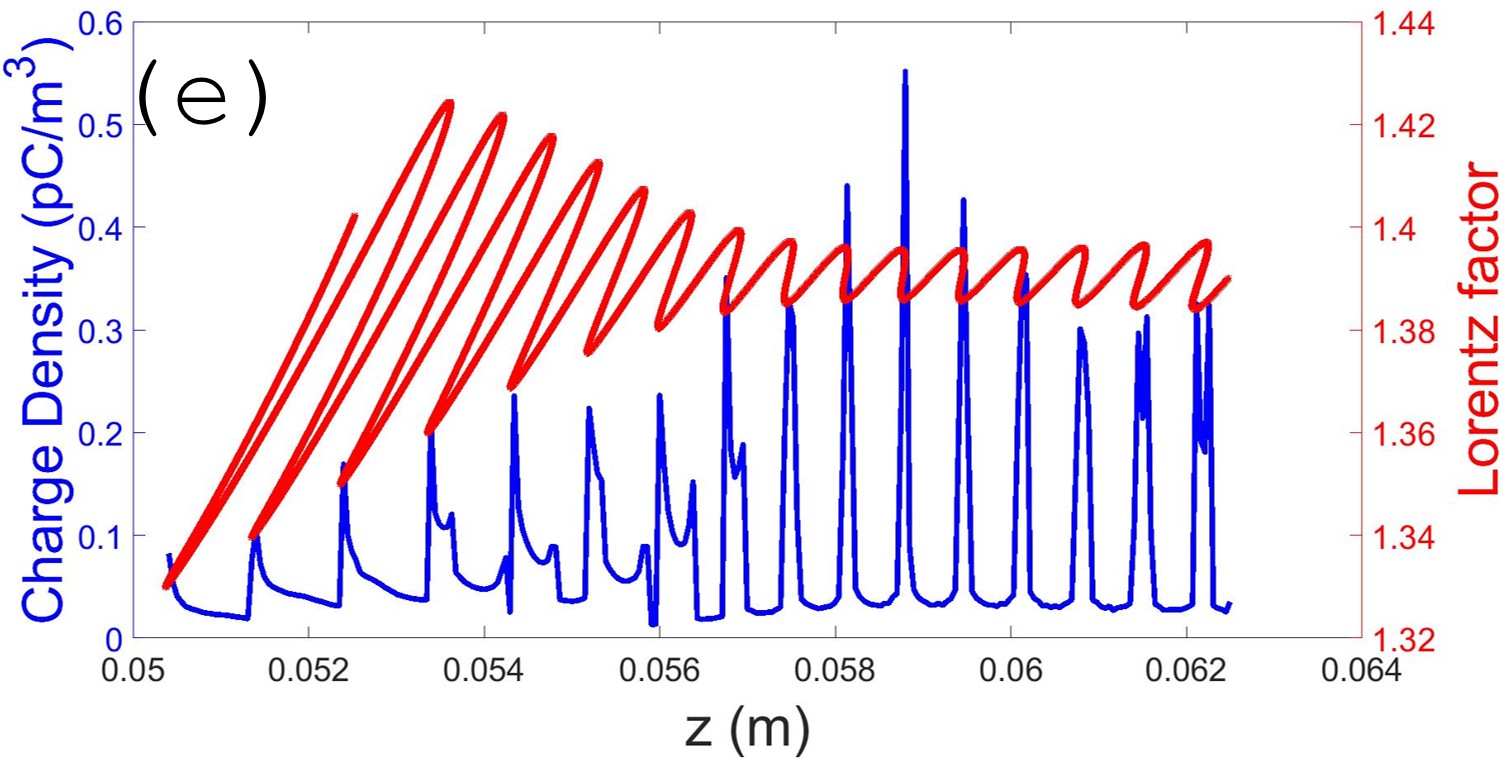}\hspace{0.5cm}\includegraphics[height=4cm]{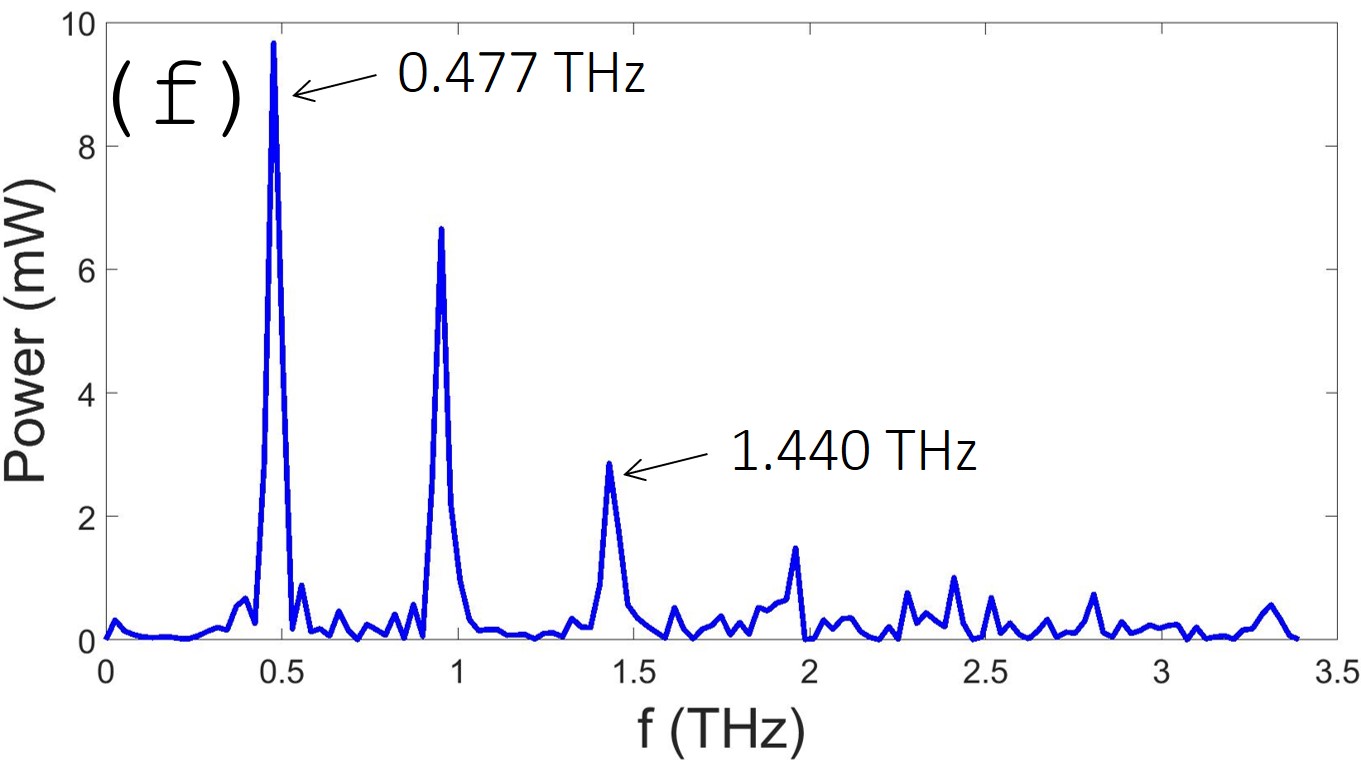}
	\caption{Comparison of phase spaces 2 cm after the DLW for (a) 1\%, (c) 3\%, (e) 2\% and frequency spectrum for (b) 1\%, (d) 3\%, (f) 2\%}\label{F:4}
\end{figure*}

The choice of the DLW geometry, as outlined in Table~\ref{T:1}, was optimized to allow for a TM$_\text{02}$ frequency be greater than 1 THz. From practical point of view, having TM$_\text{01}$ frequency greater than 1 THz is nearly impossible because in this case the vacuum channel radius of DLW should have become smaller than the typical beam size of a few microns, causing the beam to come in contact with the dielectric layer. This, in turn, would decrease the wakefield amplitude below the 10\% threshold which would require an increase in the charge of the beam, which again, in turn, would inflate the beam size.

As illustrated in Fig.~\ref{F:4}, after a certain drift distance, the phase space of the bunch transforms from a sine shaped to a sawtooth shaped (shock wave). At this point, the space charge force starts breaking the modulated but continuous beam into a train of microbunches: energy modulation results in charge density modulation. Each microbunch now contains fast- and slow-moving particles. This causes modulation in the current density where the width of this peak will determine the central frequency and the bandwidth, which can be extracted at this point by Fourier transformation. The beam compression in the rising edges of the sawtooth waveform results in the presence of higher order mode frequencies that can be resolved. As the microbunch train keeps drifting, the sawtooth waveform continuously transforms and so does the spectrum.

The following simulations use an energy band width of 1, 2 and 3 \% corresponding to a 0.5, 1 and 1.5 keV difference between each energy band to find the optimal setting. This is based on a previous application of the GPT polyenergetic wakefield model\cite{26} for undulator simulations that has shown that the energy width of an energy bin should not exceed 3\%. All snapshot comparisons were made at 0.25 ns, a position behind DLW corresponding to 2 cm. This position was found as the most optimal such that TM$_\text{01}$ carries most of energy and sawtooth shapes is such that the THz peak has narrowest band. The 2 cm number was found by executing multiple simulations around 2.5 cm benchmark. This benchmark was obtained analytically using the ballistic bunching formalism\cite{19}: it can be shown that the linear compression ratio $R_{56}$ to achieve microbunching at 200 keV is approximately --1 cm. $R_{56}$ is the negative ratio of the drift distance over relativistic $\gamma^2$ (1.6 in our case). Hence, after about 2.5 cm of drift behind DLW the perfect sawtooth like structure is expected to form.

It was found that 1\% over defines the polyenergetic Green’s function approach. As seen from Fig.~\ref{F:4}(a,b), the obtained fundamental frequency varies by almost 70 GHz, and the rest of the mode structures also significantly vary from the predicted values from the analytical calculations as reported in Table~\ref{T:2}. On the other hand, 3\% under defines our approach: due to cruder energy range partitioning it assigns higher energy to an unrealistically larger number of particles that overwhelm the resulting energy of the generated wakefield. Overestimated higher energies leads to over-acceleration such that at 0.25 ns the charge density profiles blurs out and leads to a frequency spectrum consisting of only high-frequency noise Fig.~\ref{F:4}(c,d). For 2\%, the frequency spectrum shows that the dominant mode is optimized with an output power of 10 mW, and the TM$_\text{01}$ frequency is only 7 GHz different from that calculated using the “Waveguide” code (see Fig.~\ref{F:4}(e,f)). Other TM$_\text{0m}$ modes were produced at nearly correct frequencies. There are, however, additional peaks that result from frequency splitting due to space charge, and a peak at 0.953 THz that is due to frequency mixing between the TM$_\text{01}$ and TM$_\text{02}$ (see Fig.~\ref{F:4}f). Thus, it was determined in this application that an energy band width of 2\% is optimal. The choice of 2\% over 1\% seems to be counterintuitive but the problem with 1\% over definition is caused by the discrete nature of the beam (macroparticle ensemble instead of a continuous function). It means that over definition could be caused within the sorting histogram where finer partitioning in 1\% case leads to a larger number of bins with no macroparticles present.

\begin{figure*}[!]
	\centering
	\includegraphics[width=15cm]{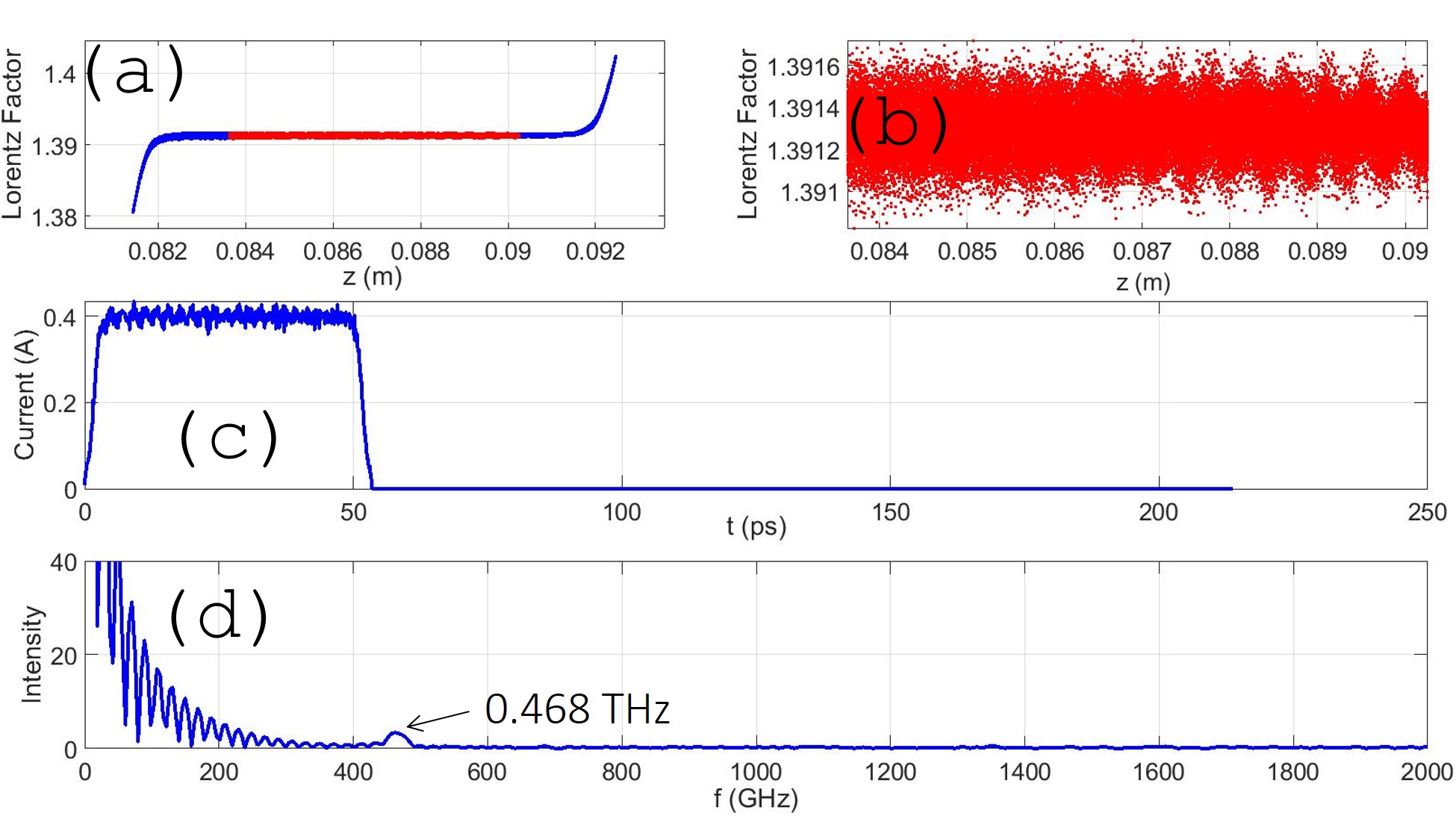}
	\caption{Comparison study from CST at transit time of $\sim$0.25 ns showing the phase space (a) and a corresponding zoomed in section showing the micro bunching and the sawtooth waveforms (b), the current density (c), the frequency spectrum where the main peak occurs at 0.468 THz (d).}\label{F:5}
\end{figure*}

A comparison with particle-in-cell (PIC) CST simulation obtained for the identical geometry further confirms the choice of 2\% energy band partitioning. As Fig.~\ref{F:5} shows, the phase space (a), zoomed in section showing the micro bunching (b), the current density (c) and the frequency spectrum (d). One difference is that CST PIC calculations show stronger energy tails at head and tail positions, resulting in the phase space resembling a propeller. However, the sawtooth waveform structure is present, and the fundamental peak does match between CST and GPT. Therefore, there is full self-consistency between analytical result, fast numerical Runge-Kutta based GPT and costly fleshed out PIC by CST.

\begin{figure*}[!]
	\centering
	\includegraphics[height=4cm]{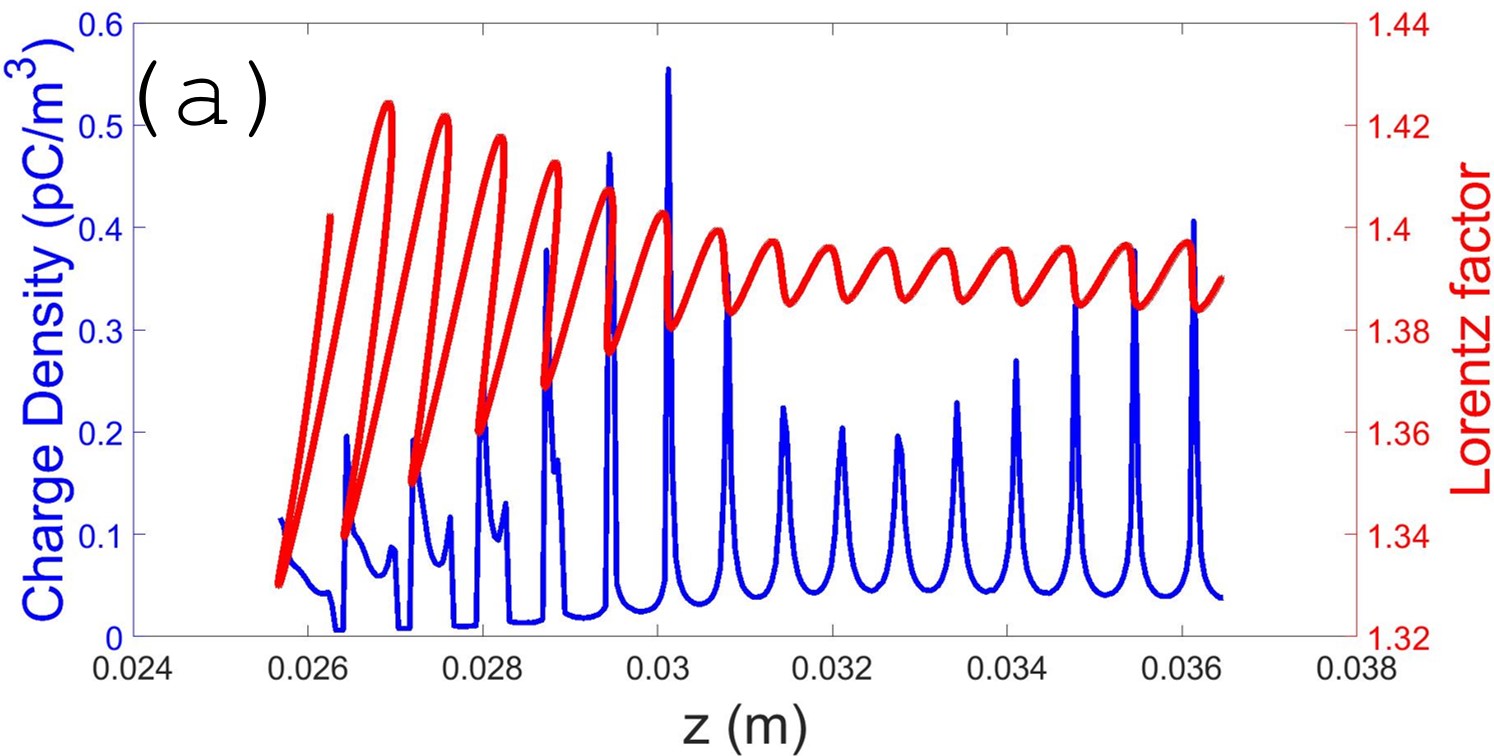}\hspace{0.5cm}\includegraphics[height=4cm]{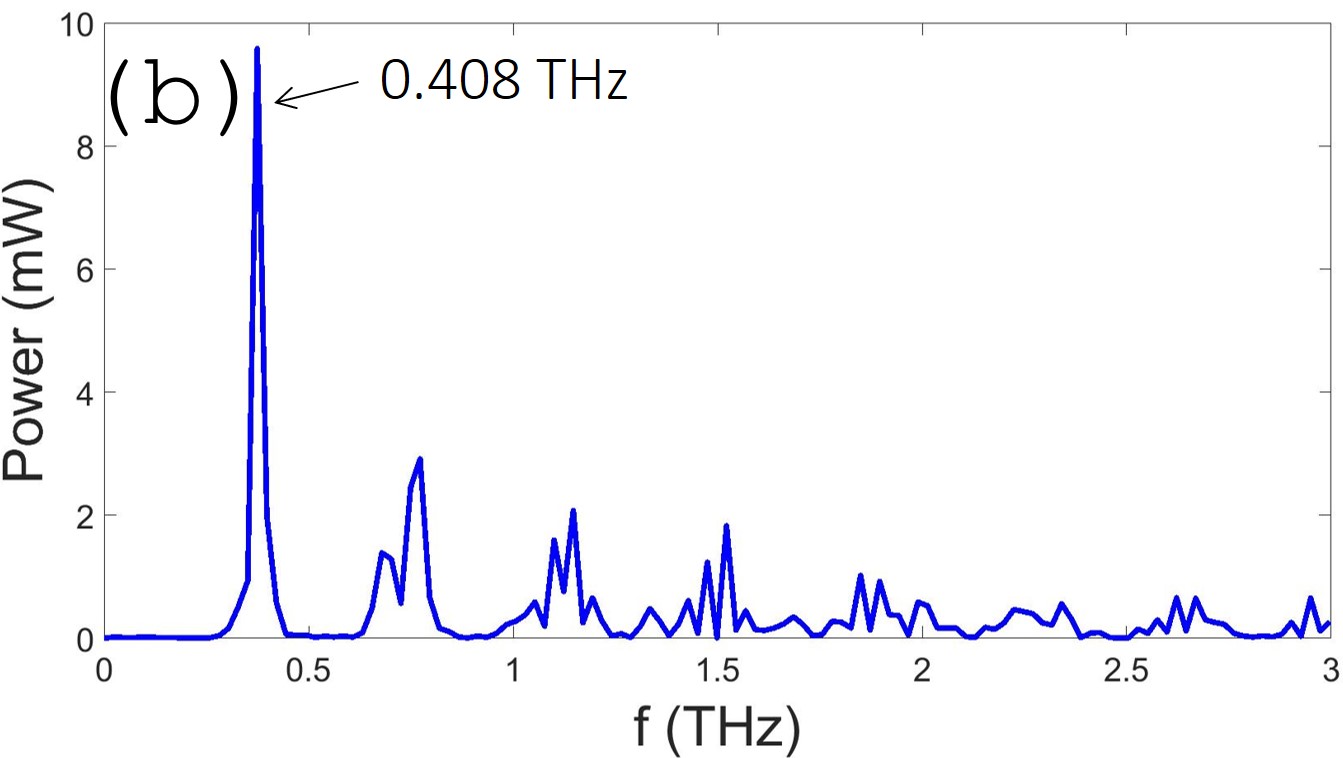}
	\includegraphics[height=4cm]{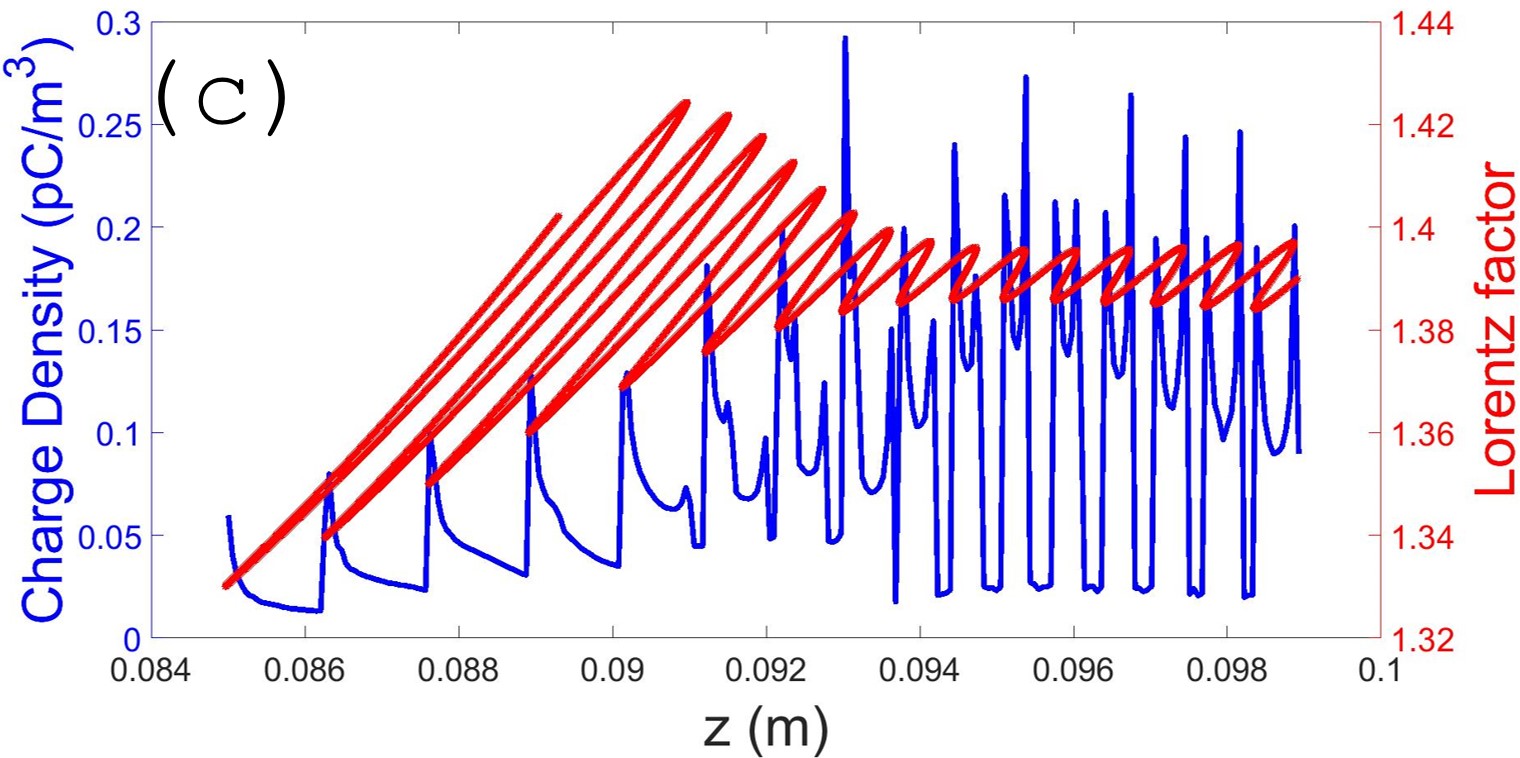}\hspace{0.5cm}\includegraphics[height=4cm]{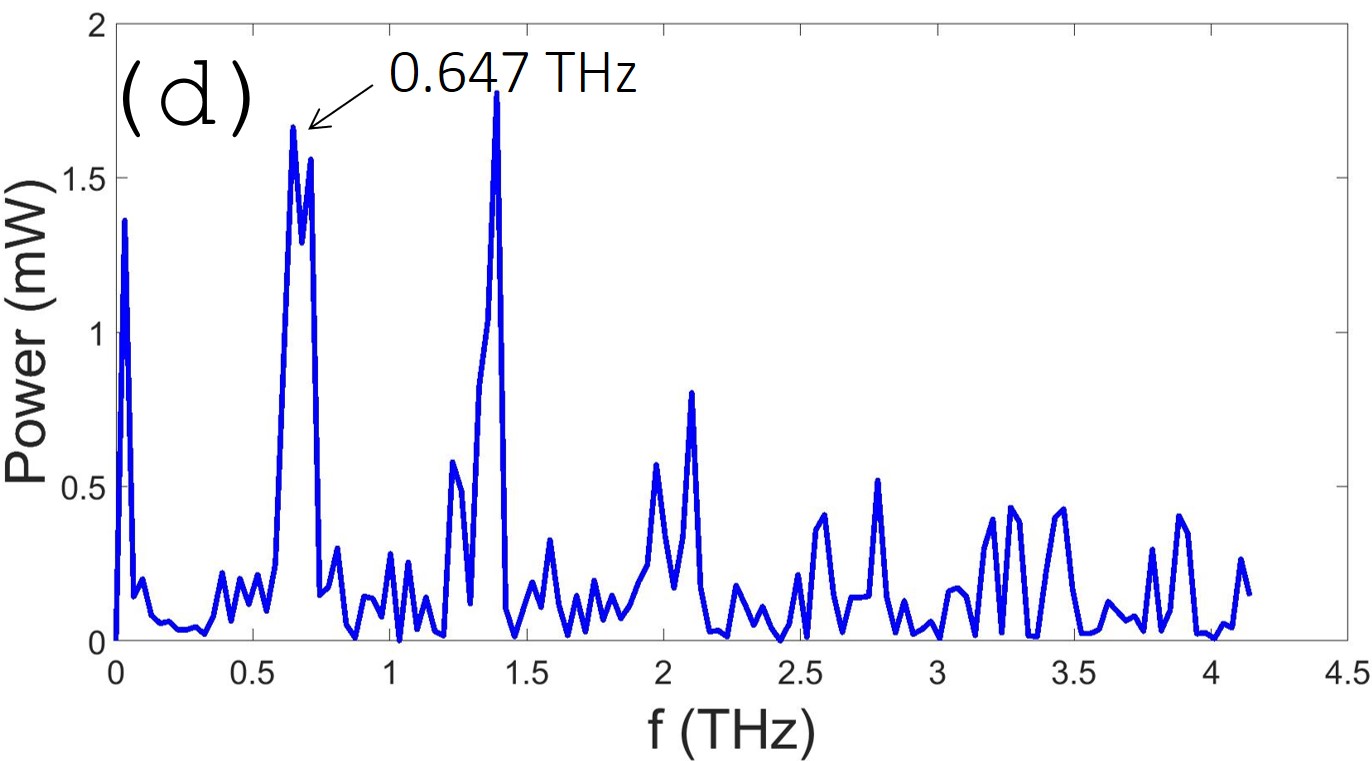}
	\includegraphics[height=4cm]{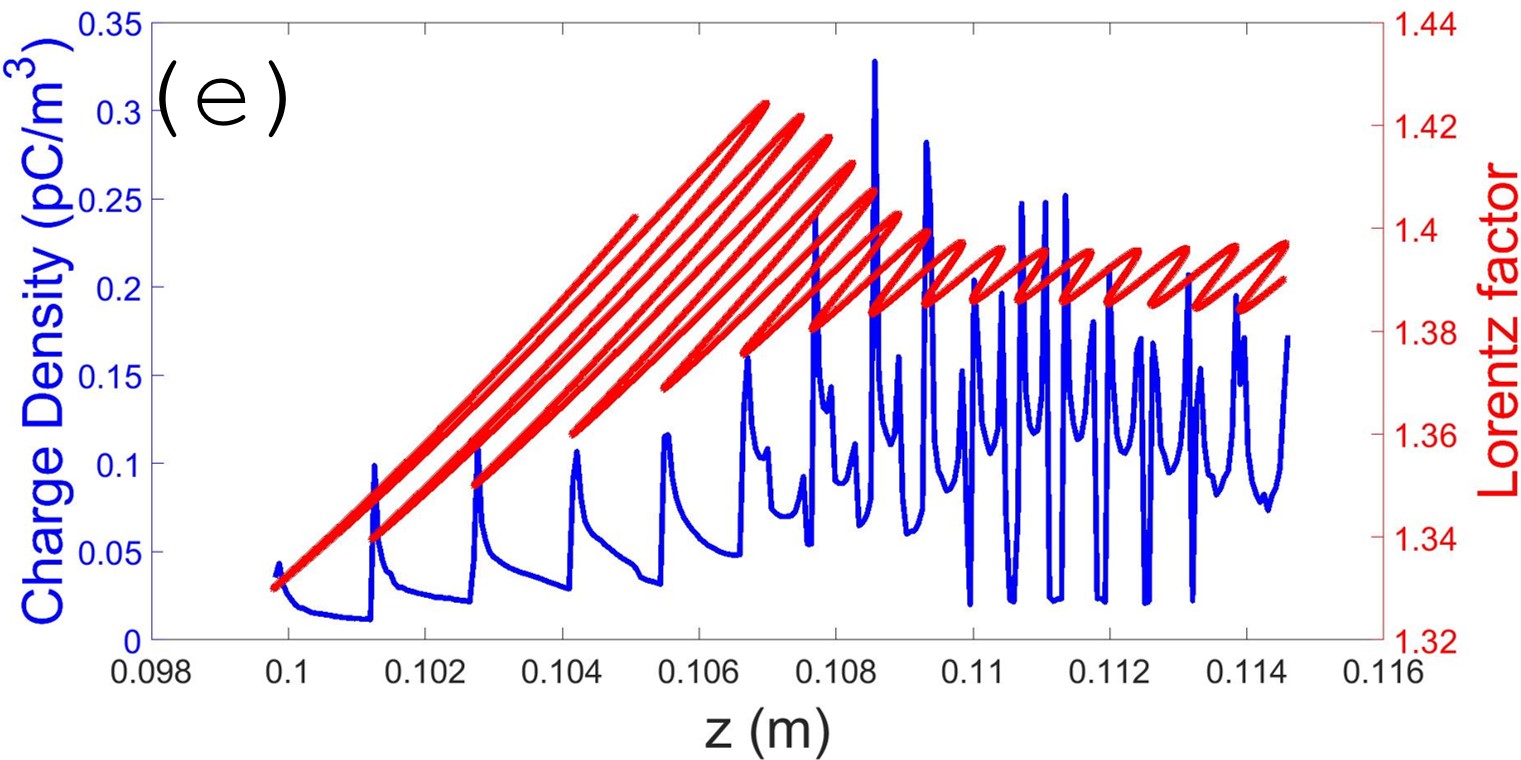}\hspace{0.5cm}\includegraphics[height=4cm]{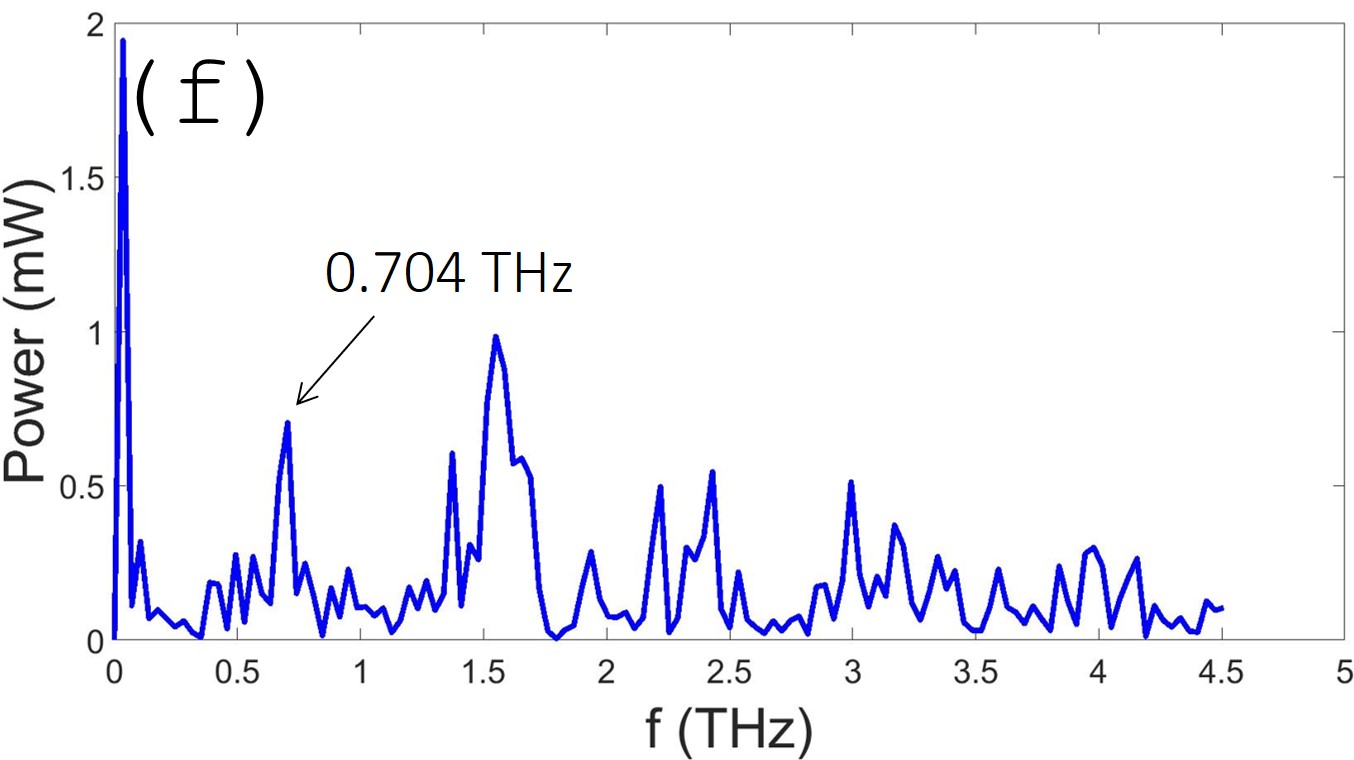}
	\caption{Evolution of the phase space/current density plots: (a) 0.125 ns, (c) 0.425 ns, (e) 0.500 ns and frequency spectra (b) 0.125 ns, (d) 0.425 ns, (f) 0.500 ns along the beam line after exiting the DLW.}\label{F:6}
\end{figure*}

\begin{figure*}[!]
	\centering
	\includegraphics[width=10cm]{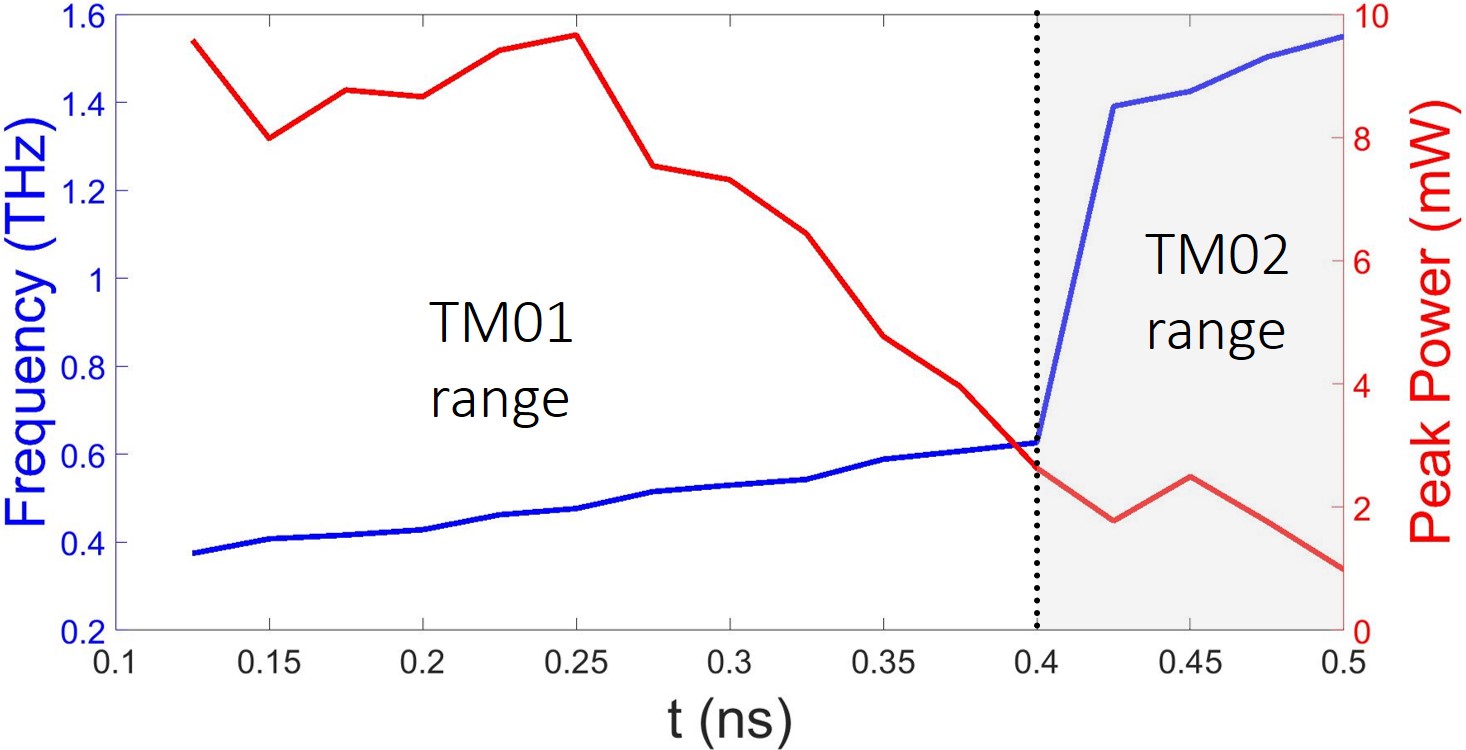}
	\caption{Tunability of the THz frequency spectrum at each snapshot after the modulated beam exited the DLW, starting at a time step of 0.125 ns and ending at 0.500 ns after which only high-frequency noise is able to be resolved. The black line shows the transition point from when the TM$_\text{02}$ mode becomes the dominant mode over the TM$_\text{01}$ mode. The peak power (red line) refers to the power of the main fundamental peak whose frequency is shown as the blue line.}\label{F:7}
\end{figure*}

More conclusions can be made if the phase and charge density space snap shots and corresponding frequency spectra are taken at different transit times. Here, the transit time is defined as the time the beam travels once it completely exits the DLW. Fig.~\ref{F:6}(a,b) shows that at 0.125 ns the sawtooth waveform did not reach the optimal and therefore it results in a lower frequency and a lower maximum power output. Then, as the beam continues to drift, the frequency is up shifted and the power continues to increase to the point where it is optimized at 0.25 ns, as painted by Fig.~\ref{F:4}(e,f). As the beam propagates, the fundamental frequency continues to increase, though the output power starts to diminish. At 0.425 ns, the dominant peak corresponds to the TM$_\text{02}$ mode, making it the dominant mode carrying the largest amount of energy as can be seen from Fig.~\ref{F:6}(c,d). At a longer drift beyond 0.425 ns, the space charge force rapidly pushes the particles further and further apart until there are only a few localized regions containing the majority of the current density. As this process continued towards 0.5 ns (Fig.~\ref{F:6}(e,f)), the beam becomes split into two micro bunches: one at the tail and one at the head due to the space charge force making a beam that would resemble that of a dumbbell. This results in no fine structure being resolved in the phase space and, subsequently, the current density resulting in a frequency spectrum consisting of only high-frequency noise; a similar picture was observed in the 3\% energy band simulation where artifact over-acceleration took place.

The proposed scheme represents a tunable system where, by simply varying the extraction distance behind the DLW, a wide variety of power-frequency spectra can be designed while maintaining the produced microwave signals narrow band. Fig.~\ref{F:7} summarizes the power gain and frequency for TM$_\text{01}$ and TM$_\text{02}$ modes as functions of the transit time (i.e. extraction distance) taken in intervals of 0.025 ns. It is apparent, the output power reduces from 9 to 3 mW as the output frequency increases from 0.37 to 0.62 THz. Here, TM$_\text{01}$ mode is dominant. When the TM$_\text{02}$ mode becomes dominant, after 0.425 ns, the output power further drops as the frequency increases from 1.39 to 1.55 THz. Additionally, it was found that the TM$_\text{02}$ mode is optimized at 0.450 ns yielding power of 2.5 mW at 1.42 THz. With a corresponding DC to AC efficiency of 6.8\%. The power gain was then found for the optimal TM$_\text{01}$ mode to be $\sim$5. The output power was $\sim$10 mW, which is higher than that produced with travelling wave tube based VEDs.

\section{Conclusion}\label{S:Conclusion}

In conclusion, it was demonstrated that nonrelativistic beams can produce high-power terahertz radiation when transmitted through a dielectrically lined waveguide when the space charge and wakefield forces are properly balanced. A novel polyenergetic Green’s function method was developed and implemented to capture and test the operation of the proposed travelling wakefield tube. It was also benchmarked against a computation making use of a particle-in-cell approach. Computations done within the polyenergetic Green’s function method allowed to create realistic simulations and demonstrated a tunability and large dynamic range of the proposed THz generator in terms of output frequency, peak power, and efficiency. Results show, the proposed 200 keV device could perform comparably when compared to the state-of-the-art equipment such as optical based THz systems. Nonrelativistic beams eliminate the need of using extremely high energy beams which would require massive amounts of space charge forces that are not trivial to produce in an experimental setting, as well as it avoids laser femtosecond seeding (pre-bunching). The proposed generator would not require extensive shielding and could become a feasible path forward to create portable/tabletop THz systems that are modular in design and that have tunable operating characteristics.

\section*{Acknowledgments}
The work by Mitchell Schneider and Ben Sims was supported by the U.S. Department of Energy Office of Science, High Energy Physics under Cooperative Agreement Award No. DE-SC0018362. The work by Emily Jevarjian and Sergey Baryshev was supported by the College of Engineering at MSU under the Global Impact Initiative. The authors would like to acknowledge helpful discussions with Dr. Stanislav Baturin (Northern Illinois University).

\bibliography{references}

\end{document}